\documentclass[preprint,amsmath,amssymb,aps,prd,showpacs]{revtex4-1}

\usepackage{epsfig}
\usepackage{dcolumn}
\usepackage{bm}
\usepackage{tikz}
\usepackage{graphicx, graphics, color}
\usepackage{dcolumn}
\usepackage{bm}
\usepackage{hyperref}
\usepackage[mathlines]{lineno}

\newcommand{\be}{\begin{equation}}
\newcommand{\ee}{\end{equation}}
\newcommand{\ben}{\begin{eqnarray}}
\newcommand{\een}{\end{eqnarray}}

\newcommand{\cO}{{\cal O}}

\newcommand{\cL}{{\cal L}}

\newcommand{\cE}{{\cal E}}
\newcommand{\cF}{{\cal F}}

\newcommand{\p}{\partial}
\newcommand{\na}{\nabla}

\newcommand{\tphi}{\tilde \phi}
\newcommand{\tpsi}{\tilde \psi}

\newcommand{\hSi}{{\hat \Sigma}}

\newcommand{\hg}{\hat g}
\newcommand{\hR}{\hat R}

\newcommand{\tG}{\tilde G}
\newcommand{\tH}{{\tilde H}}
\newcommand{\tg}{\tilde g}

\newcommand{\ep}{\epsilon}

\newcommand{\ga}{\gamma}

\newcommand{\tR}{{\tilde R}}

\newcommand{\tna}{\tilde \na}

\newcommand{\te}{\tilde e}

\begin{document}

\title{Uniqueness of electric-magnetic spacetimes with massive particle spheres}
\author{Marek Rogatko} 
\email{rogat@kft.umcs.lublin.pl}
\affiliation{Institute of Physics, 
Maria Curie-Sklodowska University, 
20-031 Lublin, pl.~Marii Curie-Sklodowskiej 1, Poland}

\date{\today}

\begin{abstract}
Uniqueness of the four-dimensional static, asymptotically flat, Einstein-Maxwell spacetime with both electric and magnetic charges,
containing non-extremal massive particle sphere, being an inner boundary in it, 
has been proved. It is isometric to Reissner-Nordstr\"om spacetime with electric/magnetic charges.
In contrast to the previous results concerning the classification of photon spheres, it describes the existence of the entire set of spacetme foliations, a set of
massive particle sphere addressed to the various energies of the particles. 
The conformal positive energy, positive mass theorem and adequate conformal transformations constitute the mail tools in the proof.
\end{abstract}

\maketitle

\section{Introduction}
In the light of the recent achievements the Event Horizon Telescope  (EHT) Collaboration in obtaining the first images of supermassive black holes 
and measuring the polarisation of light being 
the signature of magnetic fields in the vicinity of the black hole edge \cite{eht1}-\cite{eht mag5}, one can observe the growth of interests in studies 
of photon and particle orbits around compact objects, like black hole, wormholes and compact stars. Especially the region of spacetime where photon orbits are closed,
are throughly analyzed both from theoretical and observational point of view.

The concepts of  {\it photon sphere} and {\it photon surface}  \cite{vir00,cla01} attract much attention. It results from the active studies of black hole shadows and
searches for the imprints of physics beyond the Standard Model, as well as, testing the radius of black hole shadow in alternative theories of gravity, which are different than those predicted by
Einstein theory.

It turns out that the features of {\it photon spheres} resemble the properties of the black hole event horizon.
The studies of {\it photon sphere} properties reveal that it is totally umbilical hypersurface (i.e., its second fundamental form is a pure trace)
with constant mean curvature and surface gravity, being strongly resembled to black hole event horizon. On the other hand,
from black hole theory one knows that the presence of black hole event horizon enables to classify asymptotically flat spacetimes in terms of their asymptotic charges 
(authorizes the uniqueness theorems for various kind of black hole solutions).
Therefore
the concept of {\it photon sphere} can be treated as an alternative way of obtaining the classifications of black hole spacetimes (proving the uniqueness theorem for them)
 \cite{heu96}-\cite{rog24}. 

The generalizations of the uniqueness theorem for $n$-dimensional spacetime have been also under intensive explorations. Namely, in Ref. \cite{ced21} the higher-dimensional 
problem of {\it photon sphere} and uniqueness of higher-dimensional Schwarzschild spacetime was investigated. On the other hand,
the electro-vacuum $n$-dimensional case was treated in \cite{jah19}, and studies of {\it trapped photons},  in the spacetime of higher-dimensional Schwarzschild-Tangherlini black hole, have
been performed in \cite{bud20}. In Ref. \cite{rog25} the problem of uniqueness for higher dimensional electro-magnetic non-extremal solution of Einstein gravity with
$(n-2)$-form gauge fields, containing a {\it photon sphere}, has been found.

It turns out that in the case of non-spherical geometry the {\it photon sphere} is deformed into a non-spherical photon surface \cite{gib16}, or even
disappears \cite{sho17}. Some other aspects geometrical of these objects regarding also photon surfaces in stationary spacetime with rotation have been revealed
in \cite{yos17b}-\cite{kob22}.

Moreover, 
the  {\it photon sphere} and {\it surface} can also play the key role in studies of Penrose inequalities \cite{shi17}-\cite{fen20}.

Recently, the generalization of the {\it photon sphere} concept to the case of {\it massive charged particle surface/sphere} has been proposed.
They describe the case of timelike hypersurfaces to which any wordline
of particles initially touching to them remains in the hypersurface in question \cite{kob22a}-\cite{bog23a}.

In Ref. \cite{kob24} the problem of the uniqueness of static vacuum asymptotically flat spacetimes with massive particle spheres has been investigated.
In contrast to the previous theorems concerning {\it photon sphere} uniqueness, the obtained results lead to the existence of an entire spacetime foliation which is sliced by a set of massive particle spheres devoted to various energies of the particles.

In our paper we shall consider the problem of the uniqueness of static asymptotically flat 
spacetimes constituting the solution of 
Einstein-Maxwell gravity with electric $Q_{(F)}$ and magnetic $Q_{(B)}$ charges, with the line element given by
\be
ds^2 = - \Bigg( 1 -\frac{2M}{r} + \frac{Q_{(F)}^2 +Q_{(B)}^2}{r^2} \Bigg) dt^2 + \frac{dr^2}{\Big( 1 -\frac{2M}{r} + \frac{Q_{(F)}^2 +Q_{(B)}^2}{r^2} \Big) }
+ r^2 d\Omega^2,
\ee
where $d\Omega^2$ is the metric of the unit sphere, possessing a non-extremal {\it massive particle sphere} as an inner boundary.

The influence of magnetic field on the massive particle sphere region is very interesting due to the measurements and observations of black hole magnetic field
by EHT Collaboration and in the context of future planned experiments \cite{eht mag1}-\cite{eht mag5},  \cite{dal18}, as well as, anticipated next generation of EHT.

In general the trajectories of charged particles deviate from geodesics due to the Lorentz forces, however in our considerations we take into account
 a static spacetime with timelike Killing vector field, in which one has that magnetic and electric potentials are proportional to each other (see Sec. II B).

Our paper is organized as follows. In Sec. II we recall
 the basic features of Maxwell gauge field in static spacetime. Sec. III will be devoted to the basic characteristics of {\it massive particle sphere}
with electric and magnetic charges. Then
the functional dependence among lapse function and aforementioned charges has been found.
It constitutes the key ingredient for authorizing that the 
{\it massive particle sphere} has scalar constant curvature.  
In Sec. IV we elaborate the uniqueness proof by means of the conformal positive energy theorem, and alternatively by means of positive mass theorem. In both cases
the adequate conformal transformations will play key roles.
Sec. V concludes our investigations.

In order to have the correspondence with the results obtained in the case of vacuum uniqueness of {\it massive particle sphere} \cite{kob24}, we use for the anti-symmetrisation and 
symmetrisation symbols of the forms $d_{[a}d_{b]}=d_a d_b-d_bd_a$ and $d_{(a}d_{b)}= d_a d_b + d_b d_a$.

\section{Static spacetime}
The crucial part in our consideration will ordain a static spacetime and the behavior of gauge fields in this background.
In order to proceed further, let us briefly recall the basic features of Maxwell gauge field equation under the condition of the presence
of timelike Killing vector fields in the spacetime.

\subsection{Stationary Killing vector field and gauge field equations of motion}

The standard form of Einstein-Maxwell equations of motion is revealed by doing
variation of the action 
\be
S_{EM} = \int  d^4x  \sqrt{-g} \Big( R
- F_{\mu \nu} F^{\mu \nu} \Big),
\label{ac}
\ee  
with respect to $A_\mu$, 
where $g$ sets for the determinant of the four-dimensional metric tensor, while
$ F_{\mu \nu} = 2 \na_{[\mu} A_{\nu ]}$ stands for the  $U(1)$-gauge field strength. It implies 
\be
\na_{\mu} F^{\mu \nu } = 0, \qquad R_{\mu \nu} = T_{\mu \nu}(F),
\ee
On the other hand, $T_{\mu \nu} = - \delta S/ \sqrt{-g} \delta g^{\mu \nu} $, the energy momentum tensor for gauge field is given by
\be
T_{\mu \nu}({F}) = 2 F_{\mu \rho} F_{\nu}{}^{\rho} - \frac{1}{2} g_{\mu \nu} F^2.
\label{te}
 \ee
One assumes that in the elaborated  spacetime admits an asymptotically timelike Killing vector field $k_{\delta}$ and
the field strength is stationary
$\cL_k~F_{\alpha \beta} = 0.$
By virtue of the explicit form of energy momentum tensor $T_{\alpha \beta}(F)$, it also satisfies the condition $\cL_k~T_{\alpha \beta}(F) = 0$.

The existence of stationary Killing vector field $k_a$ justifies the concept of
the twist vector $\omega_a$, which implies
\be
\omega_a = \frac{1}{2}~\ep_{abcd}~k^b~\na^c~k^d,
\ee
and the fact that for any Killing vector field we obtain the relation
$\na_\alpha~\na_\beta \chi_\ga = - R_{\beta \ga \alpha}{}{}^{\delta}~\chi_\delta$, leads to the following:
 \be
\na_\beta ~\omega_\alpha = \frac{1}{2}~\ep_{\alpha \beta \ga \delta}~k^{\ga}~R^{\delta \chi}~k_{\chi}.
\label{rrq}
\ee

On this account it can be revealed that for a twist vector $\omega_\alpha$ one obtains
\be
\na_\alpha~\Big( \frac{\omega^\alpha}{N^4} \Big) = 0,
\ee
 where $N^2 = - k_\ga~k^\ga$.

The timelike Killing vector field, also allows us to define
electric and magnetic components for gauge field strengths $F_{\alpha \beta}$, which imply the following:
\be
E_{\alpha} = - F_{\alpha \beta}~k^\beta, \qquad
B_{\alpha} = \frac{1}{2}~\ep_{\alpha \beta \ga \delta}~k^\beta~F^{\ga \delta},
\ee
as well as to rewrite $F_{\alpha \delta}$ as
\be
N^2~F_{\alpha \beta}= - 
~k_{[\alpha} E_{\beta]}+ 
\ep_{\alpha \beta \ga \delta}~k^\ga~B^{\delta}.
\ee
As far as the equations of motion is concerned, they are given by
\ben \label{prop1}
\na_{\alpha}\bigg( {E^{\alpha} \over N^2} \bigg) &=& 2~{B^{\ga} \over N^4}~\omega_\ga,\\ \label{prop2}
\na_{\alpha}\bigg( {B^{\alpha} \over N^2} \bigg) &=& - 2~{E^{\ga} \over N^4}~\omega_\ga.
\een
Accordingly, by virtue of the above and relattion
$\na_{[ \ga} F_{\alpha \beta ]} = 0$, Maxwell source-free equations can be rewritten in the form as follows:
\be
\na_{[\alpha}E_{\beta ]} = 0,\qquad
\na_{[\alpha}B_{\beta ]} = 0.
\ee
Further, the assumption that
we consider the simply connected spacetime enables us to define $E_{\alpha}$ and $B_{\alpha}$ by means of 
electric and magnetic potentials given by
$E_{\alpha} = \na_{\alpha} \psi_{F},~
B_{\alpha} = \na_{\alpha} \psi_B.$
On the other hand, using the equation (\ref{rrq}), the {Poynting flux} in Einstein-Maxwell theory of gravity with electric and magnetic charges yields
\be
\na_{[\alpha} \omega_{\beta ]} = 
2~ E_{[ \alpha} B_{\beta ]}.
\label{oom}
\ee

\subsection{Static spacetime with electric/magnetic potentials}
In our paper we shall consider a smooth Riemannian manifold being 
static spacetime, with timelike Killing vector $k_\alpha$. Moreover, we
 supposes the existence of a smooth lapse function $N: M^3 \rightarrow R^{+}$, such that
$M^4 = R \times M^3$. The line element of the aforementioned spacetime is subject to the relation
\be 
ds^2 =  g_{\mu \nu} dx^\mu dx^\nu = - N^2 dt^2 + g_{ab} dx^a dx^b,
\label{metric}
\ee
where we set $N$ and $g_{ab}$ as time independent, determined on the hypersurface of constant time. In a static spacetime the timelike Killing vector
is of the form
\be
k_\alpha = N~m_\alpha,
\ee
where $m_\alpha$ constitutes a future-directed timelike unit vector and one has that $- k_\alpha k^\alpha = N^2$.

The spacetime in question is asymptotically flat, which means that it contains a data set $(\Sigma_{end},~g_{ij},~K_{ij})$ with gauge fields $A_\mu$, subject to the condition that
$\Sigma_{end}$ constitutes a manifold diffeomorphic to $R^{(3)}$ minus a 
closed 
unit ball at the origin of $R^{(3)}$. We have also 
asymptotic behaviors of $g_{ij},~F_{\mu \nu}$, provided by the following:
\ben
\mid g_{ij} &-& \delta_{ij} \mid + r~ \mid \p_a g_{ij} \mid + \dots + r^k ~\mid \p_{a_1 \dots a_k} g_{ij} \mid \\ \nonumber
&+& r \mid K_{ij} \mid
+ \dots + r^k~ \mid \p_{a_1 \dots a_k} K_{ij} \mid \le \cO \big( \frac{1}{r} \Big), \\
F_{ \alpha \beta} &+& r ~\mid \p_a F_{\alpha \beta} \mid + \dots + r^k \mid \p_{a_1 \dots a_k} F_{\alpha \beta}\mid  \le \cO \Big(\frac{1}{r^2} \Big).
\een

The Einstein-Maxwell equations of motion yield
\ben \label{1emd}
{}^{(g)}\na_i {}^{(g)}\na^i N &=& \frac{1}{N} \Big( {}^{(g)}\na_i \psi_{F} {}^{(g)}\na^i \psi_{F} + {}^{(g)}\na_i \psi_{B} {}^{(g)}\na^i \psi_{B} \Big),\\ \label{2emd}
N~{}^{(g)}\na_i {}^{(g)}\na^i \psi_{F} &=& {}^{(g)}\na_i N~{}^{(g)}\na^i \psi_{F},\\
N~{}^{(g)}\na_i {}^{(g)}\na^i \psi_{B} &=& {}^{(g)}\na_i N~{}^{(g)}\na^i \psi_{B},\\
{}^{(g)} R &=& \frac{1}{N^2} \Big( 
{}^{(g)}\na_i \psi_{F}~{}^{(g)}\na^i \psi_{F} + {}^{(g)}\na_i \psi_{B}~{}^{(g)}\na^i \psi_{B} \Big),\\ \nonumber
{}^{(g)} R_{ij} &=& \frac{1}{N} {}^{(g)}\na_i {}^{(g)}\na_j N +
\frac{1}{N^2} \Big[
g_{ij} \Big( {}^{(g)}\na_k \psi_{F} {}^{(g)}\na^k \psi_{F} + {}^{(g)}\na_k \psi_{B} {}^{(g)}\na^k \psi_{B} \Big) \\ \label{3emd}
&-& 2 ~\Big( {}^{(g)}\na_i \psi_{F} {}^{(g)}\na_j \psi_{F} + {}^{(g)}\na_i \psi_{B} {}^{(g)}\na_j \psi_{B} \Big) \Big],
\een
where one denotes ${}^{(g)}\na_i $ as the covariant derivative with respect to metric tensor $g_{ij}$, while
${}^{(g)} R_{ij}$ is the three-dimensional Ricci tensor. ${}^{(g)} R$  accounts for the Ricci scalar curvature.

The crucial point for our further studies is that in static spacetimes with Killing vector $k_\mu$ the twist vector $\omega_\alpha$ given by equation
 (\ref{oom}) is equal to zero. This fact implicate proportionality between magnetic and electric fields \cite{heu96}. However
 because of the fact that electric one-form is spacelike ($k_\mu$ is timelike), every one-form parallel and orthogonal to it vanishes (\ref{prop1})-(\ref{prop2}).
Additionally asymptotic conditions imposed on electric and magnetic potentials lead to the conclusion that
\be
\psi_B = \mu~\psi_F,
\label{elmag}
\ee
where by $\mu$ we set a constant.

\section{Geometry of massive particle surface, sphere in static asymptotically flat spacetime with electric and magnetic potentials}
By a {\it massive particle surface} \cite{kob24}
one understands a timelike hypersurface, say $S$, immersed in a spacetime manifold, for every point of which $p \in S$ and every vector $v^\alpha$
belonging to the tangent space $T_p S$, one has that $v^\alpha k_\alpha \mid_p = -\cE_k$ and $v^\alpha v_\alpha = - m^2$, and there exists 
a geodesics $\ga$ for a particle with mass $m$, energy $\cE_k$, and charge such that $\frac{d \ga}{ds} (0) = v^\alpha \mid_p$, as well as, $\ga \subset S$.
{\it Massive particle surface} is nowhere orthogonal to Killing vector field $k_\alpha$.

The above definition envisages that any geodesic attributed to a particle with energy $\cE_k$ and mass $m$ which is initially tangent
to the massive particle surface will remain tangent to it.

On the other hand, if $n^\alpha$ is normal to the massive particle surface in question, one has that the first and second fundamental forms imply \cite{kob24}
\be
h_{a b} = g_{a b} - n_a n_b, \qquad
K_{a b} = H \Big(h_{a b} + \frac{m^2}{\cE_k^2} k_a k_b \Big) + \frac{\te_A}
{\cE_k} \cF_{a b},
\label{kh}
\ee
where $K_{ab} = h^\beta_a~ h^\ga_b \na_\beta n_\ga$ and
\be
 \cF_{a b} = \frac{1}{2} n^d~F_{d (a} k_{b )},
 \ee
 which can be rewritten having in mind the features of static spacetime and the adequate definitions, as follows: 
 \be
  \cF_{a b} = \frac{1}{N^2} n^d\Big( E_d + B_d \Big)~k_a k_b.
  \ee
$H$ is a scalar function on {\it massive particle surface}, $\te_A$ is charge. 

Because of the fact that everywhere on $S$, $k^\alpha n_\alpha =0$, one can find that \cite{kob24}
\be
\cL_k n_\beta = 0, \qquad \cL_k h_{ab} = 0, \qquad
\cL_k K_{ab} = 0, \qquad \cL_k H = 0.
\ee
It can be deduced, by the direct comparison of the geometric formula which takes place
if $\bar S$ a spatial section of a surface $S$ cut by a hypersurface $\Sigma$ then the geometrical considerations \cite{yos17}, \cite{yos20}-\cite{yos20a} yield the following:
\be
K_{ab} = {\bar K}_{a b} - m_a m_b ~n^k \na_k \ln N, \qquad h_{ab} = {\bar h}_{ab} - m_a m_b,
\ee
and a comparison with the relation (\ref{kh}) reveals that
\be
{\bar K}_{a b} = H ~{\bar h}_{a b},
\label{kb}
\ee
and
\be
n^{k} \na_k \ln N = H \Big( 1 - \frac{N^2~m^2}{\cE_k^2} \Big) 
- \frac{\te_A}{\cE_k} 
~n^k \Big( E_k + B_k \Big).
\label{lnn}
\ee
Relation (\ref{kb}) exhibits the fact that a spatial section of {\it massive particle surface } is a totally umbilical one, with a spatial curvature given by ${\bar K}_a{}^a = 2 H$.

Additionally, in what follows, we suppose that spatial section is connected, closed and compact.

\subsection{Massive particle sphere}
From now on, we shall refine our considerations to the problem of {\it massive particle sphere} case. As in Refs. \cite{kob24}, by a {\it massive particle sphere} one 
will understand the case when ${}^{(g)} \na_b N =0$ on $S$. Additionally we assume the non-extremal condition for {\it massive particle sphere}
given by $1 > \frac{N^2 m^2}{\cE_k^2}$ (which stems from the equation (\ref{lnn})).

The {\it massive particle sphere} is defined as a {\it massive particle surface} with a constant lapse function $N$, the auxiliary
conditions are imposed on the electric and magnetic charges in the theory in question. 
We assume
 that the lapse function regularly foliates the manifold outside the {\it massive particle sphere}, which implies that all level
sets with $N= const$ are topological spheres. It implicates that outside {\it massive particle sphere} holds the following condition $1/\rho^2 = {}^{(g)}\na_i N {}^{(g)}\na^i N \neq 0$.

\subsection{Spatial mean curvature of massive particle sphere}

In order to analyze the basic features of spatial mean curvature, we take into account
 the Codazzi relation
\be
{}^{(g)} R _{ab} n^a Y^b = \Big( {}^{(g)} \na_b K_a {}^b - {}^{(g)} \na_a K_m{}^m \Big) Y^a,
\label{cod}
\ee
multiplied by an arbitrary tangent vector $Y^b$ to the sphere in question. The extrinsic curvature is given by the relations (\ref{kh}).

Using the fact that ${}^{(g)} \na_a (k^a k_b) =0$ \cite{kob24}, and that
electric, magnetic fields $E_a,~B_a$  ($E_a$ is normal to massive particle sphere, by its definition
and the results of subsection D,
show that this is the case for $B_a$ in static spacetime)
are normal to the massive particle sphere, the exact form of the left-hand side of (\ref{cod}), given by
\ben
{}^{(g)}R_{cd}~n^c~Y^d &=& 2 \frac{1}{N^4}~k_a~Y^a~k_b~n^b~\Big( E_m E^{m} + B_m B^{m} \Big) \\ \nonumber
&-&
2 \frac{1}{N^2} \Big( E_a E_b + B_a B_b \Big)
Y^a~ n^b + \frac{n_k Y^k}{N^2} \Big( E_m E^{m} + B_m B^{m} \Big).
\een
as well as, 
$E_a Y^a = 0,~B_a Y^a=0,~n^d k_d = 0$,
we arrive at
\be
- \Big(2 - \frac{m^2 N^2}{\cE_k^2} \Big)~{}^{(g)} \na_a H ~Y^a =0.
\ee
Because of the fact that we are interested in non-extremal sphere, the term in brackets in the above relation in non-zero, then for arbitrary tangent vector $Y^a$ one obtains that
${}^{(g)} \na_a H =0$. Thus $H$ is constant at the considered sphere.

Thus for an arbitrary vector $Y^\beta$, the mean curvature of the considered {\it massive particle sphere} is constant.\\
It can be also shown  \cite{yaz15b} that $\cL_X (n^j {}^{(g)}\na_j N) =0$, where $X$ is an arbitrary tangent vector to the sphere, envisaging that
$n^j {}^{(g)}\na_j N$ is constant on it.

On the other hand, the scalar curvature implies
\be
{}^{(g)} R = 2 H^2 \Big( 3 - \frac{2 m^2 ~N^2}{\cE_k^2} \Big) - 4 \frac{H  \te_A}
{\cE_k} \Big( n^k E_k + n^aB_a \Big) 
+ \frac{2}{N^2} \Big( n^a E_a ~n^b E_b + n^a B_a ~n^j B_j \Big).
\ee

\subsection{Scalar curvature of electric-magnetic massive particle sphere}
In order to find the scalar curvature of {\it massive particle sphere} we implement 
contracted Gauss equation provided by 
\be
{}^{(p)}R = {}^{(g)}R - 2~{}^{(g)}R_{ij}n^in^j + K_{a}{}^{a}~K_{m}{}^{m} - K_{ij}~K^{ij},
\ee
which in the case under consideration reduces to
\be
{}^{(g)}R - 2~{}^{(g)}R_{ij}n^in^j  = {}^{(p)}R - 2 H^2.
\ee
As a result we get the following:
\be
{}^{(p)} R = \frac{2}{N^2} \Big( n^a E_a ~n^b E_b + n^a B_a ~n^k B_k \Big)
+ 4 H^2 \Big( \frac{3}{2} - \frac{m^2 N^2}{\cE_k^2} \Big),
\ee
which reduces to the one obtains in \cite{kob24}, when we have no charge.

In order to show that {\it massive particle sphere} has constant a scalar curvature, 
we should to justify that $E_a n^a$ nad $B_k n^k$ are constant on the sphere.
We have pointed out
that $n^j {}^{(g)}\na_j N$ is constant on the {\it massive particle sphere} (for the proof see Ref. \cite{yaz15b}), and in the next subsection one authorizes that
electric and magnetic potentials are function of $N$, see relation (\ref{funcn}). Thus one can conclude that $E_a n^a$ nad $B_k n^k$ are constant on the sphere in question.
This fact justifies the statement that {\it massive particle sphere}
has a constant scalar curvature.

\subsection{Functional dependence - lapse function electric and magnetic potentials}
\label{func}
This subsection will be devoted to the constancy problem of $E^a n_a$ and $B_c n^c$, on {\it massive particle sphere}. Using the attitude presented by Israel \cite{isr67},
we prove the constancy of the products of electric/magnetic vectors and unit normals to the {\it massive particle sphere}.

To begin with, we introduce coordinates on $N=const,~t=const$ manifold given by
\be
g_{ab} dx^a dx^b = {}^{(2)}g_{ab} dy^a dy^b + \rho^2 dN^2,
\ee
which enables us to write the equations of motion for electric/magnetic potentials in the forms given by
\be
\frac{1}{\sqrt{{}^{(2)}g}} \frac{\p}{\p N}\Big[ \sqrt{{}^{(2)}g} \frac{\phi_F}{N} \Big] = - \frac{\Big( \rho~\psi_F^{;a} \Big)_{;a}}{N},\\
\label{in1}
\ee
where we set
\be
\frac{\p \psi_F}{\p N} = \rho~\phi_F. 
\ee
In the above one implements the relation between electric and magnetic potentials given by the relation (\ref{elmag}).

On the other hand, the gravitational relation implies
\be
\frac{1}{\rho^2} \frac{\p \rho}{\p N} = K + \frac{2 \rho (1+\mu^2)}{N}~\Big( \phi_F^2  
+ \psi_{F;a} \psi_F^{;a} 
\Big),
\label{in2}
\ee
where $K = K_m{}^m$ denotes
 the extrinsic scalar curvature of $N=const$ spacetime.
Consequently with the equations (\ref{in1})-(\ref{in2}) we can propose the integral identity written as
\ben \label{iquality}
\frac{1}{\sqrt{{}^{(2)}g}}
 \frac{\p}{\p N}  \Big[
 \sqrt{{}^{(2)}g}
  \Big( \frac{1}{N}
F(N, \tpsi ) \tphi &+& \frac{G(N, \tpsi )}{\rho} \Big) \Big] \\ \nonumber
= A~\rho~\Big( \tphi^2 + \tpsi_{; a} \tpsi^{;a} \Big) + C~\tpsi  &+& \frac{1}{\rho} \frac{\p G}{\p N}
- \frac{1}{N} \Big( F~\rho~\tpsi^{;a} \Big)_{;a},
\een
where functions $F$, $G$ are differentiable and arbitrary, while $\tpsi = \sqrt{1 + \mu^2}~ \psi_F$. 
In the definition of new potential $\tpsi$ one used the dependence of electric and magnetic potentials in the static spacetime.
For the same reason one has that $\tphi = \sqrt{1 + \mu^2}~ \phi_F$, while
the functions $A$ and $B$ are written in the forms as
\be
A = \frac{1}{N} \Big( G + \frac{\p F}{\p \tpsi} \Big) ,\qquad
B = \frac{1}{N} \frac{\p F}{\p N} + \frac{\p G}{\p \tpsi}.
\ee
The main aim is to get the integral conservation law from the relation  (\ref{iquality}), therefore we restrict to the case when
$ A = B = \frac{\p G}{\p N} = 0$. 

As was revealed in Ref. \cite{isr67} the general solutions of the above
over-determined linear system of differential equation for $F$ and $G$, will comprise a linear combination of the particular solutions, i.e.,
\be
F =1,~ G=0, \qquad F = 2 \tpsi,~ G=1, \qquad F = 2\tpsi^2 - N^2,~G = 2 \tpsi.
\label{fandg}
\ee 
The integration of equation (\ref{iquality}), with implementation of all aforementioned values of functions $F$ and $G$ given by (\ref{fandg}),
with respect to two boundary surfaces $\Sigma_0$ and $\Sigma_\infty$, and having in mind
asymptotic conditions imposed on fields given by \cite{isr67}:\\
 1) for approaching to $\Sigma_\infty$ one has that $r \psi_F  \rightarrow  Q_{(F)},~ r^2 \phi_F \rightarrow -Q_{(F)},
~\frac{\rho}{r^2} \rightarrow \frac{1}{M}$,\\
2) for $\Sigma_0$ we have that $\phi_F = \cO(N),~\psi_{F;a} = \cO(N)$,\\
3) on $\Sigma_0$ $\psi_F$ and $1/\rho$ are constant, \\
reveal the following:
\ben
&{}& \int_{\Sigma_0} dS \Big( \frac{\phi_F}{N} \Big) = - Q_{(F)},\\
2 \Big( 1 &+& \mu^2 \Big) \psi_{(0) F} \int_{\Sigma_0} dS \Big( \frac{\phi_F}{N} \Big) + \frac{S_0}{\rho_0} = M,\\
2 \Big( 1 &+& \mu^2 \Big) \psi_{(0) F}^2 \int_{\Sigma_0} dS \Big( \frac{\phi_F}{N} \Big) + 2 \frac{S_0}{\rho_0} \psi_{(0) F} = Q_{(F)},
\een
where $S_0$ denotes the area of two-space $\Sigma_0$.\\
In the derivation of the above we also use the fact that the integral 
of two-dimensional divergence over a closed $N=const$ space disappears.

The above derivation envisages the  functional dependence among $N_0$ lapse function on $\Sigma_0$, ~$\psi_{(0) F}$ electric potential at
$\Sigma_0$ and the constant $\mu$ bounded magnetic and electric potentials:
\be
 2 \Big( 1 + \mu^2 \Big) \psi_{(0) F}^2 + 2 \psi_{(0) F} \frac{M}{Q_{(F)}} - 1 = N_0^2,
\label{funcn}
 \ee
 as was mentioned above $\psi_{(0) F}$ and $N_0$ are constant on the considered hypersurface and $\psi_F \rightarrow 0$, as $r \rightarrow \infty$. 

The equations (\ref{funcn}) is valid not only on the surface in question but also in all its exterior region. Namely, let us compose the divergence identity based on the 
above equations
\be
\frac{1}{2} {}^{(g)}\na^j \Bigg[\Big(-N^2 + 2 (1+\mu^2) \psi_F^2 + \frac{2 \psi_F M}{Q_{(F)}} -1\Big) ~\xi_j \Bigg] = N~\xi_k \xi^k,
\label{gauss}
\ee
where $\xi^m$ is of the form as follows:
\be
\xi^k = - {}^{(g)} \na^k N + \frac{1}{N} \Big( 2(1+\mu^2) \psi_F~{}^{(g)} \na^k \psi_F + \frac{M}{Q_{(F)}} {}^{(g)} \na^k \psi_F \Big).
\ee
The asymptotic behaviors of $N, ~\psi_F$, and the
fact that $N>0$ in the exterior region of {\it massive particle sphere}, as well as, the application
of Gauss theorem to the relation (\ref{gauss}), conclude that $\xi^k = 0$. By taking in the above relation the value of integration constant equal to one,
one obtains the  functional dependence among electric/magnetic potentials and  $N$.
It proves the constancy of $E^a n_a$ and $B_c n^c$ on {\it massive particle sphere}, implying that ${}^{(g)} R$
is a constant scalar curvature.

Consequently, it can be observed that the presence of magnetic charge does not change the basic features of {\it massive particle sphere}.
Qualitatively the constancy of its mean curvature and scalar curvature are the same, however quantitively they are different.
These all characteristics are effected by the modified
potential $\tpsi = \sqrt{1 + \mu^2}~ \psi_F$, on which magnetic potential imprints its influence.

\subsection{Auxiliary formulae}
In this subsection we derive additional formulae describing the charge (electric/magnetic) influences 
on the {\it massive particle sphere}, for the isometric embedding $(\Sigma^2,~\sigma_{ij}) \hookrightarrow (M^3,~g_{ij})$. 

Let us commence with contracted Gauss relation, which in the case under consideration is provided by
\be
N~{}^{(\sigma)} R =  \frac{2}{N} \Big( E_a E_b + B_a B_b \Big) n^a n^b
+ 4 H~n^k~{}^{(g)}\na_k N + 2 H^2 N.
\label{kkk}
\ee
Integration  of (\ref{kkk}) over the hypersurface $\Sigma$  and applying 
the Gauss-Bonnet theorem reveal
\be
N_0 = \frac{1}{4\pi~N_0} \Big( E_a E_b + B_a B_b \Big) n^a n^b
A_\Sigma + 2 H~M_{Phs} + \frac{1}{4 \pi} H^2~A_\Sigma~N_0,
\label{no}
\ee
where we have denoted the area of the hypersurface $\Sigma$ by $A_{\Sigma} = \int_\Sigma d \Sigma$ and
the mass of the {\it massive particle sphere} by
\be
M_{phs}  = \frac{1}{4 \pi} n^k~{}^{(g)}\na_k N ~A_\Sigma.
\ee

In the next step, one elaborates the contracted 
Gauss equation ${}^{(\sigma)}R = {}^{(p)}R - 2~{}^{(p)}R_{ij} n^i n^j$,
for $(\Sigma^2,~\sigma_{ij}) \hookrightarrow (P^3,~h_{ij})$ isometric embedding, with a unit normal $n_i$.

The same procedure as above,
reveals
\be
1 = \frac{A_\Sigma H^2}{4 \pi} \Big( 3 - \frac{2 m^2 N^2}{\cE_k^2} \Big) - \frac{1}{2 \pi} H A_\Sigma~\frac{\te_A}
{\cE_k} ~\Big( n^a E_a + n^j B_j \Big)
+ \frac{1}{4 \pi} \frac{A_\Sigma}{N^2} ~\Big( n^a E_a + n^k B_k \Big). 
\ee
Defining 
electric and magnetic charges provided by 
\be
Q_{(F)}= - \frac{A_\Sigma~E_k n^k}{4 \pi~N_0}, \qquad Q_{(B)} = - \frac{A_\Sigma~B_k n^k}{4 \pi~N_0},
\ee
and combining with the relation given by
\be
H = \frac{4 \pi~M_{phs}}{\Big(1 - \frac{m^2 N^2}{\cE_k^2 } \Big) A_\Sigma ~N},
\ee
as well as, taking into account the equation (\ref{no}), lead to the expression envisaging how electric/magnetic charges influence the area $A_\Sigma$. Namely one arrives at the 
following expression:
\be
\frac{A_\Sigma}{4 \pi} = \frac{M_{phs}^2}{N^2 \Big( 1 - \frac{m^2 N^2}{\cE_k^2} \Big)}~\Big(3 - \frac{2 m^2 N^2}{\cE_k^2} \Big) + Q_{(F)}^2 +Q_{(B)}^2
- \frac{\te_A ~
( Q_{(F)}+ Q_{(B)})~M_{phs}}{\cE_k~\Big( 1 - \frac{m^2 N^2}{\cE_k^2} \Big)}.
\ee

\section{Uniqueness of static electric-magnetic massive particle sphere}

\subsection{Conformal positive energy theorem}

This subsection will be devoted to the problem of the uniqueness of {\it massive particle sphere} with electric/magnetic charges.
In our considerations we shall use the method which was widely applied in black hole classifications  \cite{gib02}-\cite{rog22},
as well as, four and higher dimensional photon  spheres uniqueness.
The method in question is based on the implementation of conformal positive energy theorem \cite{sim99}.

The basic concept underlying the conformal positive energy theorem is to consider two asymptotically flat Riemannian $(n-1)$-dimensional manifolds with
the metric tensors connected with a conformal transformation
${}^{(\Psi)} g_{ab} = \Omega^2 ~{}^{(\Phi)} g_{ab}$, where $\Omega$ stands for the conformal factor. Moreover one has the additional relation bounded with the manifold masses
${}^{(\Psi)} m + \beta {}^{(\Phi)} m \ge 0$, under the auxiliary conditions imposed on their Ricci scalar curvature tensors ${}^{(\Psi)} R + \beta {}^{(\Phi)} R \ge 0$. 
It happens that the equality holds if and only the considered manifolds are flat.

In our proof we shall implement several conformal transformations. The first two are used in order to obtain regular hypersurfaces, on which total gravitational mass vanishes, while
the next ones were implemented in order to apply the conformal positive energy
theorem and to envisage that the static slice is conformally flat.

The last applied conformal transformation reveals that the conformal flat spacetime can be rewritten in a form showing that Einstein-Maxwell 
equations of motion reduce to Laplace equation on three-dimensional Euclidean manifold. This fact
enables one to conclude that the embedding of {\it massive particle sphere} is totally umbilical and hyperspherical, which means that each component of 
the massive particle sphere is a geometric sphere of a certain radius. The embedding in question is also rigid, i.e., one can always, without loss of generality, locate one component of 
{|it massive particle sphere} at
a certain point in the hypersurface.

To begin with  conformal transformation of the form 
$\tilde g_{ij} = N^{2} g_{ij},$
leading to the conformally rescaled Ricci tensor provided by
\be
\tR_{ij}(\tg) = \frac{2}{N^2} {}^{(g)}\na_i N~{}^{(g)}\na_j N - \frac{2}{N^2}\Big( {}^{(g)}\na_i \psi_{F} {}^{(g)}\na_j \psi_{F} + {}^{(g)}\na_i \psi_{B} {}^{(g)}\na_j \psi_{B} \Big) 
\ee
In the next step, 
one defines the quantities for electric potential $\psi_F$
\ben \label{p1}
\Phi_{1}&=& \frac{1}{2} \Big( N + \frac{1}{N} - \frac{2}{N}~ \psi_{F}^2 \Big),\\
\Phi_0 &=& \frac{\sqrt{2}}{N} ~\psi_{F},\\
\Phi_{-1} &=& \frac{1}{2} \Big( N - \frac{1}{N} - \frac{2}{N}~ \psi_{F}^2 \Big),
\een
and magnetic $\psi_B$ potential
\ben
\Psi_{1}&=& \frac{1}{2} \Big( N + \frac{1}{N} - \frac{2}{N} \psi_{B}^2 \Big),\\
\Psi_0 &=& \frac{\sqrt{2}}{N} \psi_{B},\\ \label{ps3}
\Psi_{-1} &=& \frac{1}{2} \Big( N - \frac{1}{N} - \frac{2}{N} \psi_{B}^2 \Big).
\een

It can be shown that defining the metric tensor as $\eta_{AB} = diag(1, -1, -1)$, we arrive at the following auxiliary relations:
that
\be
\Phi_A \Phi^A = \Psi_A \Psi^A = -1.
\label{ffpp}
\ee
where we denote $A = - 1, 0, 1$. Further the other symmetric tensors can be constructed, respectfully for the potential $\Phi_A$
\be 
\tG_{ij} = \tna_{i} \Phi_{-1} \tna_{j} \Phi_{-1} - \tna_{i} \Phi_{0} \tna_{j} \Phi_{0} - \tna_{i} \Phi_{1} \tna_{j} \Phi_{1},
\label{g1}
\ee
and for he potential $\Psi_{A}$
\be
\tH_{ij} = \tna_{i} \Psi_{-1} \tna_{j} \Psi_{-1} - \tna_i \Psi_{0} \tna_j \Psi_0- \tna_{i} \Psi_{1} \tna_{j} \Psi_{1},
\label{h1}
\ee
where $\tna_{i}$ stands for the covariant derivative with respect to the conformally rescaled metric $\tg_{ij}$. On the other hand,
the equation (\ref{ffpp}) reveals that
\be
\tna^{2}\Phi_{A} = \tG_{i}{}{}^{i} \Phi_{A}, \qquad
\tna^{2} \Psi_{A} = \tH_{i}{}{}^{i} \Psi_{A},
\label{ppff}
\ee
and the Ricci curvature tensor $\tR_{ij}$ connected with conformally rescaled metric $\tg_{ij}$
may be rewritten in terms of $\tG_{ij}$ and $\tH_{ij}$
\be
\tR_{ij} =  \tG_{ij} + \tH_{ij}.
\label{rr}
\ee
As far as the relations (\ref{ppff}) and (\ref{rr}) are concerned, in Refs. \cite{mar02, hoe76, sim92}, it was envisaged that 
they can be derived by varying the Lagrangian density of the form as follows:
\be
\cL = \sqrt{-\tg} \Big( \tG_{i}{}^{i} + \tH_{i}{}^{i} + \frac{ \tna^i  \Phi_{A} \tna_i  \Phi^{A}}{ \Phi_{A} \Phi^{A}} + \frac{ \tna^i  \Psi_{A} \tna_i  \Psi^{A}}{ \Psi_{A} \Psi^{A}} \Big),
\ee
where the variation procedure is conducted with respect to $\tg_{ij},~ \Phi_{A}, ~ \Psi_{A}$, and  with use of the constraint relations (\ref{ffpp}).

In order to fulfil requirements of
the conformal positive energy theorem, one introduces the other
conformal transformations, which imply
\be
{}^{(\Phi)}g_{ij}^{\pm} = {}^{(\Phi)}\omega_{\pm}^{2}~ \tg_{ij},
\qquad
{}^{(\Psi)}g_{ij}^{\pm} = {}^{(\Psi)}\omega_{\pm}^{2}~ \tg_{ij},
\ee
where the conformal factors imply
\be
{}^{(\Phi)}\omega_{\pm} = {\Phi_{1} \pm 1 \over 2}, \qquad
{}^{(\Psi)}\omega_{\pm} = {\Psi_{1} \pm 1 \over 2}.
\label{pf}
\ee
They are crucial for the construction presented, e.g., in Ref.  \cite{mas92}, in order to build
manifolds $(\Sigma_{+}^{\Phi}, {}^{\Phi}g_{ij}^{+})$,
$(\Sigma_{-}^{\Phi}, {}^{\Phi}g_{ij}^{-})$, $(\Sigma_{+}^{\Psi}, {}^{\Psi}g_{ij}^{+})$, $(\Sigma_{-}^{\Psi}, {}^{\Psi}g_{ij}^{+})$,
which can be pasted
$(\Sigma_{\pm}^{\Phi}, {}^{\Phi}g_{ij}^{\pm})$ and 
$(\Sigma_{\pm}^{\Psi}, {}^{\Psi}g_{ij}^{\pm})$ across shared minimal boundaries. The construction in question leads to the
regular hypersurfaces $\Sigma^{\Phi} = \Sigma_{+}^{\Phi} \cup \Sigma_{-}^{\Phi}$ and $\Sigma^{\Psi} = \Sigma_{+}^{\Psi} \cup \Sigma_{-}^{\Psi} $. 

The next step will be connected with checking if that total
gravitational mass on hypersurfaces $\Sigma^{\Phi}$ and $\Sigma^{\Psi}$ 
is equal to zero. In order to answer this question one implements the conformal positive energy theorem and defines 
another conformal transformation given by
\be
{\hat g}^{\pm}_{ij} = \bigg[ \bigg( {}^{(\Phi)}\omega_{\pm} \bigg)^2
 \bigg( {}^{(\Psi)}\omega_{\pm} \bigg)^{2} \bigg]^{1 \over 2}\tg_{ij},
\ee
leading to the Ricci curvature tensor
\ben \label{ric}
\hat R_\pm &=& \bigg[ {}^{(\Phi)}\omega_{\pm}^2~ {}^{(\Psi)}\omega_{\pm}^{2 } \bigg]
^{-{1 \over 2}}
\bigg( {}^{(\Phi)}\omega_{\pm}^{2} {}^{(\Phi)}R_\pm +
{}^{(\Psi)}\omega_{\pm}^{2} {}^{(\Psi)}R_\pm \bigg) \\ \nonumber
&+& 
\bigg( \hat \na _{i} \ln {}^{(\Phi)}\omega_{\pm} - {\hat \na} _{i} \ln {}^{(\Psi)}\omega_{\pm} \bigg)  
\bigg( \hat \na ^{i} \ln {}^{(\Phi)}\omega_{\pm} - {\hat \na}^{i} \ln {}^{(\Psi)}\omega_{\pm} \bigg).  
\een
The direct tedious calculations unveiled that the first term on the right-hand side of the above equation could be cast as follows:
\ben \label{pos1}
{}^{(\Phi)}\omega_{\pm}^{2}~ {}^{(\Phi)}R_\pm + {}^{(\Psi)}\omega_{\pm}^{2}~ {}^{(\Psi)}R_\pm &=& 
2~\mid {\Phi_{0} \tna_{i} \Phi_{-1}
- \Phi_{-1} \tna_{i} \Phi_{0} \over
\Phi_{1} \pm 1 } \mid^2 \\ \nonumber
&+& 2~\mid { \Psi_{0} \tna_{i} \Psi_{-1}
- \Psi_{-1} \tna_{i} \Psi_{0} \over
\Psi_{1} \pm 1} \mid^2.
\een
Thus one can conclude, that in the light of the relations (\ref{ric}) and (\ref{pos1}),  the Ricci scalar $\hat R_\pm $ is greater or equal to zero.

Having in mind
 the conformal energy theorem, it has been revealed that
$(\Sigma^{\Phi}, {}^{\Phi}g_{ij})$, $(\Sigma^{\Psi}, {}^{\Psi}g_{ij})$ and
$(\hSi, \hg_{ij})$ are flat, which implies
that the conformal factors satisfy
${}^{\Phi}\omega = {}^{\Psi}\omega$ and $\Phi_{1} = \Psi_{1}$. Moreover, we get that
$\Phi_{0} = const~ \Phi_{-1}$ and $\Psi_{0} = const~ \Psi_{-1}$. 

All the above lead to the conclusion that $(\Sigma, g_{ij})$ is conformally flat manifolds. This fact enables us to rewrite $\hg_{ij}$ in a 
conformally flat form \cite{gib02,gib02a}, given by the following:
\be
\hg_{ij} = {\cal U}^{4 } {}^{\Phi}g_{ij},
\label{gg}
\ee
with ${\cal U}$ equals to  $({}^{\Phi}\omega_{\pm} N)^{-1/2}$.

On the other hand, as far as the Ricci scalar $\hR$ is concerned, its value is equal to zero. It implicates that
the considered equations of motion can be reduced
to the Laplace equation on three-dimensional Euclidean manifold. Namely
$
\na_{i}\na^{i}{\cal U} = 0,
$
where $\na$ denotes the connection on a flat manifold. Consequently, we can define a local coordinate system, with the line element provided by
\be
{}^{\Phi}g_{ij} dx^{i}dx^{j} = {\tilde \rho}^{2} d{\cal U}^2 + {\tilde h}_{AB}dx^{A}dx^{B}.
\ee
The massive particle sphere will be located at some constant value of $\cal U$, with a radius described by a fixed value of $\rho$-coordinate \cite{gib02a}.

Thus having all the above in mind, we can define on hypersurface $\Sigma$ the metric line element as follows:
\be
\hg_{ij}dx^{i}dx^{j} = \rho^2 dN^2 + h_{AB}dx^{A}dx^{B}.
\ee

In other words, the embedding of the photon sphere into Euclidean three-dimensional space is totally umbilical,
which yields \cite{kob69}
that such embedding is hyperspherical, i.e., each component of the massive particle sphere will be a geometric sphere of a certain radius.
It happens
that the studied embedding is rigid \cite{kob69}, in the sense that we can always find one connected component of the sphere
at some fixed radius, without loss of generality.
On the other hand, if we take into account massive particle sphere at fixed radius, we have a boundary conditions of Dirichlet type for $\na_{i}\na^{i}{\cal U} = 0$. It reveals
that such solution must be spherically symmetric. 

In the last step of the proof
let us suppose that ${\cal U}_{1}$ and ${\cal U}_{2}$ are two solutions of the above Laplace equation
subject to the same of the boundary value problem and regularity, use
the Green identity and integrate over the volume element. We arrive at the following:
\be
\bigg( \int_{r \rightarrow \infty} - \int_{S} \bigg) 
\bigg( {\cal U}_{1} - {\cal U}_{2} \bigg) {\p \over \p r}
\bigg( {\cal U}_{1} - {\cal U}_{2} \bigg) d\Sigma = \int_{\Omega}
\mid \na \bigg( {\cal U}_{1} - {\cal U}_{2} \bigg) \mid^{2} d\Omega.
\ee
Having in mind the imposed boundary conditions, 
the surface integral vanishes indicating that
the volume integral is identically equal to zero. It
 leads to the conclusion that the two discussed
solutions of Laplace equation subject to the Dirichlet boundary conditions are the same.

\subsection{Positive mass theorem}
The use of the conformal positive energy theorem for the uniqueness proof is not the only way to achieve the result.
For the completeness 
one ought to mention the other energy orientated proof based on the another conformal transformation and application of positive energy theorem  \cite{mas92}, 
\cite{posen}-\cite{bun87}. 
In this attitude one looks for the conformal transformation
which enable to paste two copies of $\Sigma_\pm$ along the boundary, and consider the conformal transformations
on each copy of $\Sigma$, i.e., $\Omega_\pm^2 g_{ij}$. The conformal factors yield \cite{heu96,heu94}
\be
\Omega_{\pm} = \frac{1}{4} \Big[ \Big(1 \pm N \Big)^2 - Z Z^* \Big],
\ee
while Ricci curvature for the metric $\Omega^2 g_{ij}$ yields
\ben \label{ricci}
\frac{1}{2} \Omega^4 N^2 ~R(\Omega^2 g_{ij}) &=& \mid \Big( \Omega - N \frac{\p \Omega}{\p N} \Big) {}^{(g)}\na_i Z - 2 N \frac{\p \Omega}{\p Z^*}
{}^{(g)}\na_i N \mid^2\\ \nonumber
&-& \frac{1}{16} N^2 \mid Z {}^{(g)}\na_i Z^* - Z^* {}^{(g)} \na_i Z \mid^2.
\een
 where $Z = - \psi_F + i \psi_B$, while for the brevity of notation we set $ \Omega = \Omega_\pm$.

The relation between electric and magnetic potentials in the static spacetime assures 
that the last term in (\ref{ricci})) is equal to zero and it leads to the conclusion that
 $(\Sigma, ~\Omega^2 g_{ij}) $ is an asymptotically flat complete three-dimensional manifold with non-negative scalar curvature
and vanishing mass. On the other hand, the use of positive energy theorem implies that the spacetime is isometric to $(R^3,~\delta_{ij})$.

However, the requirements for the positive energy theorem to be satisfied,
point out that it cannot be implemented for $(\Sigma_+,~\Omega_+^2 g_{ij})$ \cite{bun87}, but rather for
$$(\Sigma, ~g_{ij}) = (\Sigma_+,~\Omega_+^2 g_{ij}) \cup (\Sigma_- \cup \{p\},~\Omega_-^2 g_{ij}), $$
where $\{p\}$ is a point at infinity at $\Sigma_-$ \cite{bun87, mas92}. 
The conformal flatness of $(\Sigma, ~g_{ij})$ entails its spherical symmetry \cite{bun87, heu96}.

 The arguments, presented for instance in \cite{heu96,bun87,mas92,heu94}, lead to the final conclusion that the metric $g_{ij}$ is spherically symmetric
and we arrive at the uniqueness of massive particle sphere as an inner boundary
in the spacetime characterized by ADM mass $M$, electric and magnetic charges, being isometric to Reissner-Nordstr\"om spacetime.

Now we can formulate the main result of our considerations.\\
\noindent
{\bf Theorem}:\\
Let us assume that $(M^{3}, ~g_{ab},~N, ~\psi_F,~\psi_B)$, is the asymptotic flat, static, non-extremal,  
 Einstein-Maxwell electric-magnetic black hole spacetime, characterized by ADM mass, total electric charge $Q_{(F)}$, total magnetic charge $Q_{(B)}$. 
The spacetime in question possesses non-extremal {\it massive particle sphere}, being an inner boundary of it. The lapse function $N$ regularly foliates the considered manifold.
Then, the spacetime is isometric to the outer region of the {\it massive particle sphere} in the considered manifold.

The close inspection regarding the admissible parameter ranges for {\it massive particle sphere}, in the electric-magnetic static spacetime, has been conducted in
Ref. \cite{kob22}.

\section{Conclusions}
In our paper we have elaborated the uniqueness of black hole {\it massive particle sphere}
in Einstein-Maxwell gravity with electric and magnetic charges.
The special features of electric and magnetic fields in static spacetime with
asymptotically timelike Killing vector field, and the functional dependence among lapse function 
and electric, magnetic potentials, authorize that the Ricci curvature scalar of massive particle sphere is constant.

Applying
the conformal positive energy and positive energy theorems 
allow us to justify that static asymptotically flat spacetime being the solution of Einstein-Maxwell gravity with electric/magnetic charges, admitting a {\it massive particle sphere}
is isometric to static spherically Reissner-Nordstr\"om spacetime with electric and magnetic charges.

By contrast with the previous results connected with {\it photon spheres} classification,
now in the case of {\it massive particle spheres}, we obtain the set of spacetime foliations bounded with various possible energies of the particles.

\acknowledgments 
MR was partially supported by Grant No. 2022/45/B/ST2/00013 of the National Science Center, Poland.






\begin{thebibliography}{99}

%
\def\cmp#1#2#3#4{\emph{#4}, \emph{ Commun. Math. Phys.} {\bf #1} #2 (#3)}
\def\lmp#1#2#3#4{\emph{#4}, \emph{ Lett. Math. Phys.} {\bf #1} #2 (#3) }
\def\hpa#1#2#3#4{\emph{#4}, \emph{ Hell. Phys. Acta} {\bf #1} #2 (#3) }
\def\grg#1#2#3#4{\emph{#4}, \emph{ Gen. Rel. Grav.} {\bf #1} #2 (#3) }


\def\pr#1#2#3#4{\emph{#4}, \emph{ Phys. Rev.} {\bf #1} #2 (#3)}
\def\prl#1#2#3#4{\emph{#4}, \emph{ Phys. Rev. Lett.} {\bf #1} #2 (#3)}
\def\prd#1#2#3#4{\emph{#4}, \emph{ Phys. Rev. D} {\bf #1} #2 (#3)}
\def\pl#1#2#3#4{\emph{#4}, \emph{ Phys. Lett.} {\bf #1} #2 (#3) }
\def\pla#1#2#3#4{\emph{#4}, \emph{ Phys. Lett. A} {\bf #1} #2 (#3)}
\def\plb#1#2#3#4{\emph{#4}, \emph{ Phys. Lett. B} {\bf #1} #2 (#3)}
\def\prep#1#2#3#4{\emph{#4}, \emph{ Phys. Reports} {\bf #1} #2 (#3) }
\def\phys#1#2#3#4{\emph{#4}, \emph{ Physica} {\bf #1} #2  (#3) }
\def\jcp#1#2#3#4{\emph{#4}, \emph{ J. Comput. Phys.} {\bf #1} #2 (#3) }
\def\jmp#1#2#3#4{\emph{#4}, \emph{ J. Math. Phys.} {\bf #1} #2 (#3) }
\def\jpm#1#2#3#4{\emph{#4}, \emph{ J. Phys. A: Math. Gen.} {\bf #1} #2 (#3) }
\def\cpr#1#2#3#4{\emph{#4}, \emph{ Computer Phys. Rept.} {\bf #1} #2 (#3) }
\def\cqg#1#2#3#4{\emph{#4}, \emph{ Class. Quant. Grav.} {\bf #1} #2 (#3)}
\def\cma#1#2#3#4{\emph{#4}, \emph{ Computers Math. Applic.} {\bf #1} #2  (#3)}
\def\mc#1#2#3#4{\emph{#4}, \emph{ Math. Compt.} {\bf #1} #2 (#3)}
\def\apj#1#2#3#4{\emph{#4}, \emph{ Astrophys. J.} {\bf #1} #2 (#3) }
\def\apjs#1#2#3#4{\emph{#4}, \emph{ Astrophys. J. Suppl.} {\bf #1} #2 (#3)}
\def\apjl#1#2#3#4{\emph{#4}, \emph{ Astrophys. J. Lett.} {\bf #1} #2 (#3) }
\def\acta#1#2#3#4{\emph{#4}, \emph{ Acta Astronomica} {\bf #1} #2 (#3) }
\def\apl#1#2#3#4{\emph{#4}, \emph{ Ann. Physik. (Leipzig)} {\bf #1} #2 (#3) }
\def\amjp#1#2#3#4{\emph{#4}, \emph{Am. J. Phys.} {\bf #1} (#3) #2}
\def\anp#1#2#3#4{\emph{#4}, \emph{ Ann. Phys.} {\bf #1} (#3) #2}
\def\sa#1#2#3#4{\emph{#4}, \emph{ Sov. Astro.} {\bf #1} (#3) #2}
\def\sia#1#2#3#4{\emph{#4}, \emph{ SIAM J. Sci. Statist. Comput.} {\bf #1} (#3) #2}
\def\aa#1#2#3#4{\emph{#4}, \emph{ Astron. Astrophys.} {\bf #1} #2 (#3)}
\def\mnras#1#2#3#4{\emph{#4}, \emph{ Mon. Not. R. Astr. Soc.} {\bf #1} (#3) #2}
\def\npb#1#2#3#4{\emph{#4}, \emph{ Nucl. Phys. B} {\bf #1} (#3) #2}
\def\prsla#1#2#3#4{\emph{#4}, \emph{ Proc. R. Soc. London, Ser. A} {\bf #1} (#3) #2}
\def\jhep#1#2#3#4{\emph{#4}, \emph{ JHEP} {\bf #1} (#2) #3}
\def\jcap#1#2#3#4{\emph{#4}, \emph{ J. Cosmol. Astropart. Phys.} {\bf #1} #2 (#3)}
\def\nuc#1#2#3#4{\emph{#4}, \emph{ Nuovo Cimento B } {\bf #1} (#3) #2}
\def\ijmp#1#2#3#4{\emph{#4}, \emph{ Int. J. Mod. Phys. D} {\bf #1} (#3) #2}
\def\atmp#1#2#3#4{\emph{#4}, \emph{ Adv. Theor. Math. Phys.} {\bf #1} #2 (#3) }
\def\ptps#1#2#3#4{\emph{#4}, \emph{ Prog. Theor. Phys. Suppl.} {\bf #1} #2 (#3)}
\def\ptep#1#2#3#4{\emph{#4}, \emph{ Prog. Theor. Exp. Phys.} {\bf #1} #2 (#3)}
\def\lmp#1#2#3#4{\emph{#4}, \emph{ Lett. Math. Phys.} {\bf #1} #2 (#3)}
\def\cpam#1#2#3#4{\emph{#4}, \emph{ Comm. Pure Appl. Math.}  {\bf #1} #2 (#3) }
\def\adv#1#2#3#4{\emph{#4}, \emph{ Adv. Phys.}  {\bf #1} (#3) #2}
\def\zh#1#2#3#4{\emph{#4}, \emph{ Zh. Eksp. Teor. Fiz.}  {\bf #1} (#3) #2}

\def\jams#1#2#3#4{\emph{#4}, \emph{ J. Austral. Math. Soc. B} {\bf #1} (#3) #2}
\def\appa#1#2#3#4{\emph{#4}, \emph{ Acta Phys. Polonica A} {\bf #1}, (#3) #2}
\def\nat#1#2#3#4{\emph{#4}, \emph{Nature} {\bf #1}, (#3) #2}
\def\science#1#2#3#4{\emph{#4}, \emph{Science} {\bf #1}, (#3) #2}
\def\arcmp#1#2#3#4{\emph{#4}, \emph{Annual Rev. of Cond. Matter Physics} {\bf #1}, (#3) #2}
\def\jcap#1#2#3#4{\emph{#4}, \emph{JCAP} {\bf #1}, (#3) #2}
\def\conphy#1#2#3#4{\emph{#4}, \emph{Contemporary Physics} {\bf #1}, (#3) #2}
\def\ptp#1#2#3#4{\emph{#4}, \emph{ Prog. Theor. Phys.} {\bf #1} #2 (#3)}
\def\ptpexp#1#2#3#4{\emph{#4}, \emph{ Prog. Theor. Exp. Phys.} {\bf #1} #2 (#3)}

\def\apjsup#1#2#3#4{\emph{#4}, \emph{ Astrophys. J. Suppl. Ser.} {\bf #1} (#3) #2}
\def\mplb#1#2#3#4{\emph{#4}, \emph{ Mod. Phys. Lett. B} {\bf #1} (#3) #2}
\def\ijmpd#1#2#3#4{\emph{#4}, \emph{ Int. J. Mod. Phys. D} {\bf #1} (#3) #2}

\def\gravcos#1#2#3#4{\emph{#4}, \emph{ Grav. Cosmol.} {\bf #1} #2 (#3)}
\def\ajp#1#2#3#4{\emph{#4}, \emph{ Am. J. Phys.} {\bf #1} (#3) #2}
\def\appb#1#2#3#4{\emph{#4}, \emph{ Acta Phys. Polon. B} {\bf #1} (#3) #2}
%
\def\hepph#1#2{{ hep-ph }{#1} (#2)}
\def\hepth#1#2{{ hep-th }{#1} (#2)}
\def\arx#1#2{{ arXiv~}{#1} (#2)}
\def\astroph#1#2{{ astro-ph }{#1} (#2)}
\def\grqc#1#2{{ gr-qc }{#1} (#2)}
\def\ibid#1#2#3#4{\emph{#4}, {\it ibid.} {\bf #1} #2 (#3)}
\def\cag#1#2#3#4{\emph{#4}, \emph{ Commun. Anal. Geom.} {\bf #1} #2 (#3) }
\def\contmath#1#2#3#4{\emph{#4}, \emph{ Contemp. Math.} {\bf #1} #2 (#3)}
\def\epjc#1#2#3#4{\emph{#4}, \emph{ Eur. Phys. J. C} {\bf #1} #2 (#3) }
\def\revphys#1#2#3#4{\emph{#4}, \emph{Reviews in Phys.} {\bf #1} #2 (#3) }
\def\science#1#2#3#4{\emph{#4}, \emph{ Science} {\bf #1} #2 (#3) }

%
\bibitem{eht1}
K. Akiyama et al., \apjl{875}{L1}{2019}{First M87 Event Horizon Telescope Results. I. The Shadow of the Supermassive Black Hole}.
\bibitem{eht2}
K. Akiyama et al., \apjl{930}{L12}{2022}{First Sagitarius A* Event Horizon Telescope Results.1. The shadow of the supermassive black hole in the center of the Milky Way}.

\bibitem{eht mag1}
K. Akiyama et al., \apjl{910}{L12}{2021}{First M87 Event Horizon Telescope Results. VII. Polarization of the Ring}.
\bibitem{eht mag2}
K. Akiyama et al., \apjl{910}{L13}{2021}{First M87 Event Horizon Telescope Results. VIII. Magnetic Field Structure near The Event Horizon}.
\bibitem{eht mag3}
K. Akiyama et al., \apjl{957}{L20}{2023}{First M87 Event Horizon Telescope Results. IX. Detection of Near-horizon Circular Polarization}.

\bibitem{eht mag4}
K. Akiyama et al., \apjl{964}{L25}{2024}{First Sagittarius A*
Event Horizon Telescope Results. VII. Polarization of the Ring}.
\bibitem{eht mag5}
K. Akiyama et al., \apjl{964}{L26}{2024}{First Sagittarius A*
Event Horizon Telescope Results. VIII. Physical Interpretation of the Polarized Ring}.




\bibitem{vir00}
K.S. Virbhadra and G.F.R. Ellis, \prd{62}{084003}{2000}{Schwarzschild black hole lensing}.
\bibitem{cla01}
C-M. Claudel, K.S. Virbhadra, and G.F.R. Ellis, \jmp{42}{818}{2001}{The geometry of photon surfaces}.


\bibitem{heu96}
M. Heusler, {\it Black Hole Uniqueness Theorems}, 
Cambridge: Cambridge University Press, 1996.

\bibitem{ced15a}
C. Cederbaum, \contmath{667}{86}{2015}{Uniqueness of photon spheres in static vacuum asymptotically flat spacetimes}.

\bibitem{yaz15}
S. Yazadjiev, \prd{91}{123013}{2015}{Uniqueness of the static spacetimes with a photon sphere in Einstein-scalar field theory}.
\bibitem{yaz15b}
S. Yazadjiev and B. Lazov, \cqg{32}{165021}{2015}{Uniqueness of the static Einstein-Maxwell spacetimes with a photon sphere}.


\bibitem{ced15}
C. Cederbaum and G. Galloway, \cag{25}{303}{2017}{Uniqueness of photon spheres via positive mass rigidity}.
\bibitem{ced16}
C. Cederbaum and G. Galloway, \cqg{33}{075006}{2016}{Uniqueness of photon spheres in electro-vacuum spacetimes}.

\bibitem{rog16}
M. Rogatko, \prd{93}{064003}{2016}{Uniqueness of photon sphere for Einstein-Maxwell-dilaton black holes with arbitrary coupling constant}.

\bibitem{yaz16}
S. Yazadjiev and B. Lazov, \prd{93}{064003}{2016}{Classification of the static and asymptotically flat Einstein-Maxwell-dilaton spacetimes with a photon sphere}.

\bibitem{tom17}
Y. Tomikawa, T. Shiromizu, and K. Izumi, \ptpexp{2017}{033E03}{2017}{On the uniqueness of the static black hole with conformal scalar hair}.
\bibitem{tom17b}
Y. Tomikawa, T. Shiromizu, and K. Izumi, \cqg{34}{15504}{2017}{On uniqueness of static spacetimes with non-trivial conformal scalar field}.
\bibitem{yos17}
H. Yoshino, \prd{95}{044047}{2017}{Uniqueness of static photon surfaces: Perturbative approach}.

\bibitem{kog20}
Y. Koga, \prd{101}{104022}{2020}{Photon surfaces as pure tension shells: Uniqueness of thin shell wormholes}.
\bibitem{yaz21}
S. Yazadjiev, \prd{104}{124070}{2021}{Classification of static asymptotically flat spacetimes with a photon sphere in Einstein-multiple-scalar field theory}.
\bibitem{rog24}
M. Rogatko, \prd{109}{024056}{2024}{Uniqueness of photon sphere for Reissner-Nordstr\"om electric-magnetic system}.





\bibitem{ced21}
C. Cederbaum and G. Galloway, \jmp{62}{032504}{2021}{Photon surfaces with equipotential time slices}.
\bibitem{jah19}
S. Jahns, \cqg{36}{235019}{2019}{Photon sphere uniqueness in higher-dimensional electrovacuum spacetimes}.
\bibitem{bud20}
M. Budgen, \cqg{37}{015001}{2020}{Trapped photons in Schwarzschild-Tangherlini spacetimes}
\bibitem{rog25}
M. Rogatko, \prd{112}{044052}{2025}{Higher-dimensional electric-magnetic photon sphere uniqueness}.

\bibitem{gib16}
G. W. Gibbons and C. M. Warnick, \plb{763}{169}{2016}{Aspherical photon and anti-photon surfaces}.
\bibitem{sho17}
A. Shoom, \prd{96}{084056}{2017}{Metamorphoses of a photon sphere}.



\bibitem{yos17b}
H. Yoshino, K. Izumi, T. Shiromizu, and Y. Tomikawa, \ptpexp{2017}{063E0}{2017}{Extension of photon surfaces and their area: Static and stationary spacetimes}.
\bibitem{gal19}
D. V. Gal'tsov and K. V. Kobialko, \prd{99}{084043}{2019}{Completing characterization of photon orbits in Kerr and Kerr-Newman metrics}.

\bibitem{ced19}
C. Cederbaum and S. Jahns, \grg{51}{79}{2019}{Geometry and topology of the Kerr photon region in the phase space}, ~\ibid{51}{154(E)}{2019}{Correction to: Geometry and topology 
of the Kerr photon region in the phase space}.
\bibitem{gal19b}
D. V. Gal'tsov and K. V. Kobialko, \prd{100}{104005}{2019}{Photon trapping in static axially symmetric spacetime}.

\bibitem{cao21}
L. -M. Cao and Y. Song, \epjc{81}{714}{2021}{Quasi-local photon surfaces in general spherically symmetric spacetimes}.

\bibitem{yos20}
H. Yoshino, K. Izumi, T. Shiromizu, and Y. Tomikawa, \ptpexp{2020}{023E02}{2020}{Transversely trapping surfaces: Dynamical version }.
\bibitem{kob20}
K. V. Kobialko and D. V. Gal'tsov, \epjc{80}{527}{2020}{Photon regions and umbilic conditions in stationary axisymmetric spacetimes}.

\bibitem{kog21}
Y. Koga, T. Igata, and K. Nakashi, \prd{103}{044003}{2021}{Photon surfaces in less symmetric spacetimes}.


\bibitem{kob21}
K. V. Kobialko, I. Bogush, and D. V. Gal'tsov, \prd{104}{044009}{2021}{Killing tensors and photon surfaces in foliated spacetimes}.
\bibitem{kob22}
K. V. Kobialko, I. Bogush, and D. V.  Gal'tsov, \prd{106}{024006}{2022}{Slice-reducible conformal Killing tensors, photon surfaces, and shadows}.



\bibitem{shi17}
T. Shiromizu, Y. Tomikawa, K. Izumi, and H. Yoshino, \ptpexp{2017}{033E01}{2017}{Area bound for surface in a strong gravity}.
\bibitem{yan20}
R. Q. Yang and H. Lu, \epjc{80}{949}{2020}{Universal bounds on the size of a black hole}.
\bibitem{fen20}
X. H. Feng and H. Lu, \epjc{80}{551}{2020}{On the size of rotating black holes}.



\bibitem{kob22a}
K. V. Kobialko, I. Bogush, and D. V.  Gal'tsov, \prd{106}{084032}{2022}{Geometry of massive particle surfaces}.
\bibitem{bog23}
I. Bogush, K. Kobialko, and D. V. Gal'tsov, \prd{108}{044070}{2023}{Massive particle surfaces}.
\bibitem{bog23a}
I. Bogush,K. V. Kobialko, and D. V.  Gal'tsov, \arx{2306.12888}{2023}
{\it Glued massive particle surfaces}.
\bibitem{kob24}
K. Kobialko, I. Bogush, and D. Gal'ltsov, \prd{110}{044059}{2024}{Uniqueness of the static vacuum asymptotically flat spacetimes with massive particle spheres}.




\bibitem{dal18}
Y. Dallilar et al. \science{358}{1299}{2017}{A precise measurement of the magnetic field in the corona of the black hole binary V404 Cygni}.



bibitem{yos17}
H. Yoshino, K. Izumi, T. Shiromizu, and Y. Tomikawa, \ptep{2017}{063E01}{2017}{Extension of photon surfaces and their area: static and stationary spacetimes}.
\bibitem{yos20}
H. Yoshino, K. Izumi, T. Shiromizu, and Y. Tomikawa, \ptep{2020}{023E02}{2020}{Transversely trapping surfaces: dynamical version}.
\bibitem{yos20a}
H. Yoshino, K. Izumi, T. Shiromizu, and Y. Tomikawa, \ptep{2020}{023E02}{2020}{Formation of dynamically transversely trapping surfaces and the stretched hoop cojecture}.









\bibitem{isr67}
W. Israel, \cmp{8}{245}{1967}{Event horizons in static electrovac space-time}.




\bibitem{sim99}
W. Simon, \lmp{50}{275}{1999}{Conformal Positive Mass Theorems}.



\bibitem{mar02}
M. Mars and W. Simon, \atmp{6}{279}{2002}{On uniqueness of static Einstein-Maxwell-dilaton black holes}.
\bibitem{gib02}
G.W. Gibbons, D. Ida, and T. Shiromizu, \prl{89}{041101}{2002}{Uniqueness and nonuniquess of static black holes in higher dimensions}.
\bibitem{gib02a}
G.W. Gibbons, D. Ida, and T. Shiromizu, \prd{66}{044010}{2002}{Uniqueness of (dilatonic) charged black holes and black p-branes in higher dimensions}.
\bibitem{rog02}
M. Rogatko, \cqg{19}{L151}{2002}{Uniqueness theorem for static black hole solutions of $\sigma$-models in higher dimensions}.
\bibitem{rog03}
M. Rogatko, \prd{67}{084025}{2003}{Uniqueness theorem of static degenerate and nondegenerate charged black holes in higher dimensions}.
\bibitem{rog04}
M. Rogatko, \prd{70}{044023}{2004}{Uniqueness theorem for generalized Maxwell electric and magnetic black holes in higher dimensions}.
 \bibitem{rog06}
 M. Rogatko, \prd{73}{124027}{2006}{Classification of static charged black holes in higher dimensions}.
\bibitem{rog13}
M. Rogatko, \prd{88}{024051}{2013}{Uniqueness of charged static asymptotically flat black holes in dynamical Chern-Simons gravity}.
\bibitem{rog22}
M. Rogatko, \prd{105}{104021}{2022}{Classification of static black holes in Einstein phantom-dilaton Maxwell--anti-Maxwell gravity systems}.














\bibitem{hoe76}
C. Hoenselaers, \ptp{55}{46}{1976}{Multipole moments of electrostatic space-times}.
\bibitem{sim92}
W. Simon, \cqg{9}{241}{1992}{Radiative Einstein-Maxwell spacetimes and 'no-hair' theorems}.

\bibitem{mas92}
A. K. M. Masood-ul-Alam, \cqg{9}{L53}{1992}{Uniqueness proof of static charged black hole revisited}.



\bibitem{kob69}
S. Kobayashi and K. Nomizu, {\it Foundations of Differential Geometry}, (Intersciene, New York, 1969), vol.II.







\bibitem{posen}
R. Schoen and S.T. Yau, \cmp{65}{45}{1979}{On the proof of the positive mass conjecture in general relativity},\\
E. Witten, \cmp{80}{381}{1981}{A new proof of the positive energy theorem}.
\bibitem{heu94}
M. Heusler, \cqg{11}{L49}{1994}{On the uniqueness of Reissner-Nordstr\"om solution with electric and magnetic charge}.
\bibitem{bun87}
G. L. Bunting and A. K. M. Masood-ul-Alam, \grg{19}{147}{1986}{Nonexistence of multiple black holes in asymptotically Euclidean static vacuum space-time}.










\end{thebibliography}
\end{document}